\documentclass[conference,letterpaper]{IEEEtran}

\usepackage[utf8]{inputenc} 
\usepackage[T1]{fontenc}
\usepackage{url}
\usepackage{ifthen}
\usepackage[cmex10]{amsmath} 

\usepackage{amssymb} 
\usepackage{mathtools} 
\usepackage{amsfonts} 
\usepackage{accents} 

\usepackage{mleftright}\mleftright 

\usepackage{multirow}

\usepackage[dvips]{graphicx} 
\graphicspath{{./}}
\DeclareGraphicsExtensions{.eps}

\usepackage[dvipsnames,table]{xcolor} 

\usepackage{bbm} 
\usepackage{xfrac} 
\usepackage{cancel} 
\usepackage{wrapfig} 
\usepackage{tensor} 

\usepackage{theorem}
\usepackage[style=ieee,url=false,doi=false,isbn=false]{biblatex}
\usepackage{url} 

\usepackage{psfrag} 

\usepackage{tikz}
\usetikzlibrary{positioning}

\usepackage[inline]{enumitem} 

\usepackage{array} 

\usepackage{aliascnt} 

\usepackage[draft]{hyperref} 

\usepackage{cleveref} 

\usepackage{orcidlink} 

\usepackage[ruled,algoruled,vlined]{algorithm2e}

\SetCommentSty{mycommfont}

\SetKwInput{KwInput}{Input}                
\SetKwInput{KwOutput}{Output}              
\SetKwRepeat{Do}{do}{while}%
\SetKwProg{Init}{init}{}{}

\usepackage{xparse} 

\usepackage{ifthen} 

\usepackage{comment}

\crefname{section}{Section}{Sections}
\crefname{subsection}{Section}{Sections}
\crefname{equation}{Eq.}{Equations}
\crefname{enumi}{part}{parts}
\crefname{table}{Table}{Tables}
\crefname{figure}{Figure}{Figures}
\crefname{algocf}{Algorithm}{Algorithms}

\newtheorem{theorem}{Theorem}
\crefname{theorem}{Theorem}{Theorems}

\newaliascnt{lemma}{theorem}
\newtheorem{lemma}[lemma]{Lemma}
\aliascntresetthe{lemma}
\crefname{lemma}{Lemma}{Lemmas}

\newaliascnt{definition}{theorem}

\aliascntresetthe{definition}
\crefname{definition}{Definition}{Definitions}

\newaliascnt{corollary}{theorem}
\newtheorem{corollary}[corollary]{Corollary}
\aliascntresetthe{corollary}
\crefname{corollary}{Corollary}{Corollarys}

\newaliascnt{claim}{theorem}
\newtheorem{claim}[claim]{Claim}
\aliascntresetthe{claim}
\crefname{claim}{Claim}{Claims}

\newaliascnt{conjecture}{theorem}

\aliascntresetthe{conjecture}
\crefname{conjecture}{Conjecture}{Conjectures}

\newaliascnt{question}{theorem}

\aliascntresetthe{question}
\crefname{question}{Question}{Questions}

\newaliascnt{example}{theorem}
\newtheorem{example}[example]{Example}
\aliascntresetthe{example}
\crefname{example}{Example}{Examples}

\newaliascnt{oquestion}{theorem}

\aliascntresetthe{oquestion}
\crefname{oquestion}{Open Question}{Open Questions}

\theoremstyle{plain}
\theorembodyfont{\normalfont}

\newaliascnt{remark}{theorem}

\aliascntresetthe{remark}
\crefname{remark}{Remark}{Remark}

\newtheorem{cnstr}{Construction}

\newenvironment{construction}{\begin{cnstr}}{\hfill$\Box$\end{cnstr}}
\crefname{cnstr}{Construction}{Constructions}

\crefname{step}{Step}{Steps}

\crefname{regime}{Regime}{Regimes}

\newtheorem{myalgo}{Algorithm}

\crefname{myalgo}{Algorithm}{Algorithms}

\newcounter{enumrom}
\renewcommand{\theenumrom}{(\roman{enumrom})}

\makeatletter
\renewcommand{\@endtheorem}{\endtrivlist}
\makeatother

\makeatletter
\renewcommand{\thefigure}{{\@arabic\c@figure}}
\renewcommand{\fnum@figure}{{\bf Figure\,\thefigure}}
\makeatother

\renewcommand{\leq}{\leqslant}
\renewcommand{\geq}{\geqslant}

\newcommand{\bfa}{{\boldsymbol a}}
\newcommand{\bfb}{{\boldsymbol b}}
\newcommand{\bfc}{{\boldsymbol c}}

\newcommand{\bfu}{{\boldsymbol u}}
\newcommand{\bfv}{{\boldsymbol v}}

\newcommand{\bfx}{{\boldsymbol x}}
\newcommand{\bfy}{{\boldsymbol y}}

\newcommand{\cC}{\mathcal{C}}

\newcommand{\cE}{\mathcal{E}}

\newcommand{\cH}{\mathcal{H}}

\newcommand{\cR}{\mathcal{R}}

\newcommand{\cX}{\mathcal{X}}

\DeclarePairedDelimiter\abs{\lvert}{\rvert}
\DeclarePairedDelimiter\ceilenv{\lceil}{\rceil}

\DeclarePairedDelimiter\parenv{\lparen}{\rparen}

\DeclarePairedDelimiterX\mathset[2]{\lbrace}{\rbrace}{#1 : #2}
\DeclarePairedDelimiterX\multset[2]{\lbrace\!\!\lbrace}{\rbrace\!\!\rbrace}{#1 : #2}
\DeclarePairedDelimiterX\inner[2]{\langle}{\rangle}{#1 \mathrel{},\mathrel{} #2}
\DeclarePairedDelimiterX\condparenv[2]{(}{)}{#1 \mathrel{}\delimsize\vert\mathrel{} #2}

\DeclareDocumentCommand\norm{ o m }{
\IfNoValueTF{#1}
{\left\Vert#2\right\Vert}
{\left\Vert#2\right\Vert_{#1}}
}

\DeclareDocumentCommand\der{ o m o }{
\IfNoValueTF{#1}
{
\IfNoValueTF{#3}
{\frac{d}{d{#2}}}
{\frac{d{#3}}{d{#2}}}
}
{\parenv*{\frac{d}{d{#2}}}^{#1}\IfNoValueTF{#3}{}{#3}}
}
\DeclareDocumentCommand\partder{ o m m }{
\IfNoValueTF{#1}
{\frac{\partial{#3}}{\partial{#2}}}
{\frac{\partial^{#1}{#3}}{{\partial{#2}}^{#1}}}
}
\DeclareDocumentCommand\df{ o m o }{
d\IfNoValueTF{#1}{}{^{#1}}{#2}\IfNoValueTF{#3}{}{_{#3}}
}

\DeclareMathOperator{\tr}{tr}

\DeclareMathOperator{\wt}{wt}

\DeclarePairedDelimiter{\floor}{\lfloor}{\rfloor}

\DeclareDocumentCommand\enc{ o }{
\IfNoValueTF{#1}
{\operatorname{Enc}}
{\operatorname{Enc}_{\ref*{#1}}}
}

\DeclareDocumentCommand\dec{ o }{
\IfNoValueTF{#1}
{\operatorname{Dec}}
{\operatorname{Dec}_{\ref*{#1}}}
}

\newtheorem{propT}[theorem]{{P}roposition}
\newtheorem{defn}[theorem]{Definition}

\makeatletter
\newcommand\code[1]{%
\@ifundefined{r@#1}{%
\cC_{\operatorname*{#1}}%
}{%
\cC_{\ref*{#1}}%
}%
}
\makeatother

\bibliography{literature.bib}

\renewcommand{\tr}[2][]{\mathcal{R}_{#1}(#2)}
\newcommand{\tri}[3][]{\tr[#1]{\boldsymbol{#2}}_{#3}}

\newcommand{\del}[1]{\mathrm{D}({#1})}
\newcommand{\sticky}[2]{\mathrm{DS}({#1}; {#2})}

\newcommand{\restr}[2]{\mathrm{DR}({#1}; {#2})}

\IEEEoverridecommandlockouts
\begin{document}
\title{
Correcting a Single Deletion in Reads \\ from a Nanopore Sequencer

\thanks{
  This material is based upon work supported by the National Science Foundation under Grant No. CCF 2212437. This work has also received funding from the European Research Council (ERC) under the European Union’s Horizon 2020 research and innovation programme (Grant agreement No. 801434). It was also funded by the European Union (ERC, DNAStorage, 865630). Additionally, this project was funded by the European Union (DiDAX, 101115134). Views and opinions expressed are however those of the author(s) only and do not necessarily reflect those of the European Union or the European Research Council Executive Agency. Neither the European Union nor the granting authority can be held responsible for them. 
  The work of Yonatan~Yehezkeally was supported by the Alexander von Humboldt Foundation under a Carl Friedrich von Siemens Post-Doctoral Research Fellowship.%

  \copyright 2024 IEEE. Personal use of this material is permitted. Permission from IEEE must be obtained for all other uses, in any current or future media, including reprinting/republishing this material for advertising or promotional purposes, creating new collective works, for resale or redistribution to servers or lists, or reuse of any copyrighted component of this work in other works.
}
} 


\author{%
  \IEEEauthorblockN{Anisha Banerjee\IEEEauthorrefmark{1},
                    Yonatan Yehezkeally\IEEEauthorrefmark{1},
                    Antonia Wachter-Zeh\IEEEauthorrefmark{1},
                    and Eitan Yaakobi\IEEEauthorrefmark{2}}
  \IEEEauthorblockA{\IEEEauthorrefmark{1}%
                    Institute for Communications Engineering, Technical University of Munich (TUM), Munich, Germany}
  \IEEEauthorblockA{\IEEEauthorrefmark{2}%
                    Department of Computer Science, Technion---Israel  Institute  of  Technology, Haifa 3200003, Israel}
    \IEEEauthorblockA{Email: anisha.banerjee@tum.de, 
      yonatan.yehezkeally@tum.de, 
      antonia.wachter-zeh@tum.de,
      yaakobi@cs.technion.ac.il}
      \\[-3.6ex]

}


\maketitle


\begin{abstract}
 Owing to its several merits over other DNA sequencing technologies, nanopore sequencers hold an immense potential to revolutionize the efficiency of DNA storage systems. However, their higher error rates necessitate further research to devise practical and efficient coding schemes that would allow accurate retrieval of the data stored. Our work takes a step in this direction by adopting a simplified model of the nanopore sequencer inspired by Mao \emph{et al.}, which incorporates some of its physical aspects. This channel model can be viewed as a sliding window of length $\ell$ that passes over the incoming input sequence and produces the Hamming weight of the enclosed $\ell$ bits, while shifting by one position at each time step. The resulting $(\ell+1)$-ary vector, referred to as the $\ell$-\emph{read vector}, is susceptible to deletion errors due to imperfections inherent in the sequencing process. We establish that at least $\log n - \ell$ bits of redundancy are needed to correct a single deletion. An error-correcting code that is optimal up to an additive constant, is also proposed. Furthermore, we find that for $\ell \geq 2$, reconstruction from two distinct noisy $\ell$-read vectors can be accomplished without any redundancy, and provide a suitable reconstruction algorithm to this effect.

\end{abstract}

\section{Introduction}

Our ever-increasing data storage requirements have prompted extensive research into DNA storage, as it promises high density and unmatched longevity. While there continue to be significant efforts to improve upon existing synthesis and sequencing technologies, nanopore sequencing holds particular appeal due to better portability, ability to read longer DNA strands, and real-time analysis \cite{deamerThree2016, laszloDecoding2014, kasianowiczCharacterization1996}. The sequencing operation involves the transmigration of a DNA fragment through a microscopic pore in a lipid membrane, across which a voltage difference exists. The nucleotides in the pore at a given time instant, influence the variations in the ionic current, which are measured and fed to a basecaller that predicts the nucleotides in the examined DNA strand. Despite its strengths, certain physical aspects of the nanopore sequencer lead to various distortions in the final readout. For instance, the variations in the measured current are governed by multiple nucleotides instead of just one due to the depth of the pore, thus hinting at the presence of intersymbol interference (ISI). Moreover, the DNA strand often passes through the pore unevenly, i.e., a few nucleotides may be skipped
, or some backtracking may occur. This naturally implies deletions and duplications in the final readout, respectively.

Prior work in this area was largely aimed at either developing faithful mathematical models for the sequencer or designing error-correcting codes that incorporate such models to correct errors in the readouts efficiently. For instance, the authors of \cite{maoModels2018}
introduced a channel model that incorporates ISI, deletions, and measurement noise. Upper bounds on channel capacity were also established. The work in \cite{hulettCoding2021} adopted a more deterministic model, devised an algorithm to compute the capacity of the same, and also suggested efficient coding schemes. A finite-state semi-Markov channel (FSMC)-based model, introduced more recently in \cite{mcbainFiniteState2022a}, encapsulates the effects of ISI, duplications, and noisy measurements that affect the final sequencing output. {Another promising line of work aims to design codes such that the current reading of the constituent DNA sequences, as produced by a nanopore sequencer, can be decoded accurately with high probability despite sample duplications and amplitude noise \cite{vidalConcatenatedNanoporeDNA2024, vidalErrorBoundsDecoding2023, vidalUnionBoundGeneralized2023}. 
In \cite{banerjeeErrorCorrectingCodesNanopore2024}, a specific model inspired by \cite{maoModels2018, hulettCoding2021} was considered, and an optimal single-substitution-correcting code was presented.} 

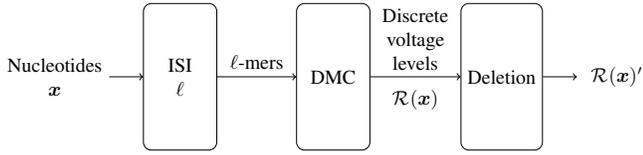
\begin{figure}[t]
	\scalebox{0.75}{
	\begin{tikzpicture}
	
	\node[align=center] (start) {Nucleotides \\ $\bfx$};
	
	\node[draw, minimum width=1.3cm,
	minimum height=2.6cm, right=0.6cm of start, rounded corners, align=center] (ISI) {ISI \\ $\ell$ };
	
	\node[draw, minimum width=1.3cm,
	minimum height=2.6cm, right=14mm of ISI, rounded corners] (dmc) {DMC};
	
	\node[draw, minimum width=1.4cm,
	minimum height=2.6cm, right=16mm of dmc, rounded corners] (del) {Deletion};
	
	\node[align=center, right=0.6cm of del] (end) {};

	\draw[->] (start) -- (ISI);
	\draw[->] (ISI) --  node[above=0.3pt]{$\ell$-mers} (dmc);
    \draw[->] (dmc) -- node[above, align=center]{Discrete \\ voltage \\ levels} node[below=3.0pt]{$\tr{\bfx}$} (del);
 
	\draw[->] (del) -- (end) node[align=center, right=0.1pt] (h) {$\tr{\bfx}'$};
	
	\end{tikzpicture}}
	\caption{Simplified model of a nanopore sequencer }
    
	\label{fig::ch_model}
\end{figure}

This work endeavors to extend \cite{banerjeeErrorCorrectingCodesNanopore2024} by designing efficient deletion-correcting codes for nanopore sequencers. To this end, we use the channel model employed in \cite{banerjeeErrorCorrectingCodesNanopore2024}. This model 
also resembles the transverse-read channel \cite{cheeTransverseReadCodesDomainWall2023, yerushalmiCapacityWeightedRead2024}, which is relevant to racetrack memories. More specifically, nanopore sequencing is interpreted as a concatenation of three channels, as illustrated in Fig.~\ref{fig::ch_model}. The ISI component, parameterized by $\ell$, signifies how the measured current depends on the $\ell$ consecutive nucleotides in the pore at any given time. This stage may be viewed as a sliding window of length $\ell$ passing over an input sequence and shifting by a single position after each time step, producing a sequence of $\ell$-mers, i.e., strings of $\ell$ symbols. Subsequently, {a discrete memoryless channel converts each of the $\ell$-mers into a discrete voltage level based on a deterministic function, in our analysis the Hamming weight. In the end, the deletion channel accounts for the effect of skipping forward 
by corrupting the sequence of discrete voltage levels with deletions.}

We now state the problem more formally. For an input $\bfx \in \Sigma_2^n$, let $\tr[\ell]{\bfx}$ represent the deletion-free channel output (Definition~\ref{def::read-vec}). 
Thus, we are interested in codes that correct $t$ deletions in $\tr[\ell]{\bfx}$ as opposed to $\bfx$ itself, to guarantee the unique recovery of $\bfx$ despite ISI, followed by at most $t$ deletions. 

To summarize the main contributions of this work, we establish a lower bound on the redundancy required by a code that corrects a single deletion in $\ell$-read vectors, and suggest an instantiation of the same whose redundancy is optimal up to an additive constant. Since nanopore sequencers tend to produce multiple erroneous reads for each input strand, we also examine how leveraging this feature might help achieve a lower redundancy requirement. To this end, we find that for any $\bfx \in \Sigma_2^n$ and $\ell \geq 2$, two distinct noisy channel outputs that arise from the same input $\bfx$, suffice to recover $\bfx$ uniquely. A suitable reconstruction algorithm is also stated.

\section{Preliminaries}

\subsection{Notations and Terminology}
In the following, we let $\Sigma_q$ indicate the $q$-ary alphabet $\{0,1,\ldots,q-1\}$. 
Additionally, $[n]$ is used to denote the set $\{1,2,\ldots, n\}$. 
Element-wise modulo operation on a vector, say $\bfy \in \Sigma_q^n$, is represented as
\begin{align*}
\bfy \bmod a\triangleq\big( y_1\bmod a, y_2\bmod a, \ldots, y_n\bmod a \big).
\end{align*}

For any vector $\bfx=(x_1, \ldots, x_n)$, we refer to its substring $(x_i, x_{i+1}, \ldots, x_j)$ as $\bfx_i^j$. The Hamming weight of a vector $\bfx$ is denoted by $\wt(\bfx)$
, while the number of runs in $\bfx$, which are of length greater than or equal to some $a\geq 1$, is represented as $\rho_{\geq a}(\bfx)$. We designate the number of all runs, i.e., when $a=1$, by $\rho(\bfx)$. We also extensively use the Hamming distance, which is defined for any two vectors $\bfx, \bfy \in \Sigma_q^n$ as 
\begin{equation*}
d_H(\bfx, \bfy)=|\{i : i\in [n], x_i \neq y_i\}|.
\end{equation*}

We focus on the case of ${q=2}$ 
and in this framework, the channel output is defined as follows.
\begin{defn}
    The $\ell$-\emph{read vector} of any ${\bfx \in \Sigma_2^n}$ is of length $n+\ell-1$ over $\Sigma_{\ell+1}$, and is denoted by
\begin{equation*}
\tr[\ell]{\bfx} \triangleq (\wt(\bfx_{2-\ell}^1), \wt(\bfx_{3-\ell}^{2}), \ldots, \wt(\bfx_{n}^{n+\ell-1})), \label{def::read-vec}
\end{equation*}
where for ease of notation we simply let $x_i=0$ for any $i\not\in [n]$ (i.e., when the above definition includes indices which are either negative or greater than~$n$). 
\end{defn}

Additionally, the $i$-th element of $\tr[\ell]{\bfx}$ is denoted by $\tr[\ell]{\bfx}_i$; that is, $\tri[\ell]{x}{i}=\wt(\bfx_{i-\ell+1}^{i})$. When clear from the context, $\ell$ will be removed from the preceding notations.

\begin{example}
    The $3$-read vector of $\bfx=(1,0,1,1,0,0)$ is given by $\tr{\bfx}=(1,1,2,2,2,1,0,0)$. Its third element is $\tri{\bfx}{3} = 2$. \label{eg::read}
\end{example}

A similar model was investigated in \cite{cheeTransverseReadCodesDomainWall2023}, wherein the output sequence, termed as the transverse-read vector, is a substring of the $\ell$-read vector as defined here, for certain parameter choices. The information limit of this transverse-read channel was computed for various parameters, and several coding schemes enabling exact recovery were presented. Some error-correcting codes for the case of $\ell = 2$ were also suggested.

Next, to facilitate our analysis, some key properties of $\ell$-read vectors are stated below.

\begin{propT}[\cite{banerjeeErrorCorrectingCodesNanopore2023}]
\begin{enumerate}[label={P\arabic*}]
\item For any $\ell \geq 1$ and $\bfx \in \Sigma_2^n$, it holds that ${\sum_{i=1}^{n+\ell-1} \tri{\bfx}{i}=  \ell \cdot \wt(\bfx) }$, where the sum is performed over integers. \label{prop::adj}
\item For any $\ell \geq 1$ and $\bfx \in \Sigma_2^n$, it holds that $|\tri{\bfx}{j+1}-\tri{\bfx}{j}| \leq 1$ for all $j \in [n+\ell-2]$. \label{prop::wt_modl}
\item For any $\bfx \in \Sigma_2^n$, $\bfx$ can be uniquely and efficiently determined from the first or last $n$ elements of $\tr{\bfx} \bmod 2$. \label{prop::rec_rx} 
\end{enumerate} \label{proposition::char}
\end{propT}

The preceding definitions can be extended to the non-binary alphabet by replacing the notion of Hamming weights with compositions, as done in \cite{banerjeeErrorCorrectingCodesNanopore2024}.

\subsection{Error Model}

To suitably define what constitutes an error-correcting construction in our framework, we first let $\del{\bfu}$ refer to the set of all vectors of length $n-1$, that can be obtained by deleting one symbol from $\bfu \in \Sigma^n$, for any alphabet~$\Sigma$. Naturally, we are interested in~$\del{\tr{\bfx}}$ for some~$\bfx\in \Sigma_2^n$. A code that corrects a single deletion in $\tr{\bfx}$ can thus be defined as follows.

\begin{defn}
For $n\geq \ell$, a code $\cC\subseteq \Sigma_2^n$ is said to be a \emph{single-deletion $\ell$-read code} if for any two distinct $\bfx, \bfy \in \cC$, it holds that $\del{\tr[\ell]{\bfx}} \cap \del{\tr[\ell]{\bfy}} = \emptyset$.
\end{defn}

\section{Correcting a Single Deletion}

A useful consequence of the inherent characteristics of $\ell\text{-read}$ vectors, summarized in Proposition~\ref{proposition::char}, is that certain deletions can be corrected immediately without any redundancy, as shown in the next lemma.

\begin{lemma}
Let $\mathcal{R}'$ arise from a single deletion on $\tr{\bfx}$ for some $\bfx \in \Sigma_2^n$. If there exists some $i \in [n+\ell-3]$ such that $|\mathcal{R}'_{i+1}-\mathcal{R}'_{i}|>1$, then $\tr{\bfx}$ (and thereby also $\bfx$) can be readily recovered. \label{lem::del_g2}
\end{lemma}

\begin{IEEEproof}
From \ref{prop::adj}, we infer that the existence of an $i \in [n+\ell-3]$ such that $|\mathcal{R}'_{i+1}-\mathcal{R}'_{i}|>1$ unambiguously reveals the error location. Note that $|\mathcal{R}'_{i+1}-\mathcal{R}'_{i}| \leq 2$. Assuming $\mathcal{R}'_{i+1}-\mathcal{R}'_{i} = 2$, we observe that the only $\bfx$ for which $\mathcal{R}' \in \del{\tr{\bfx}}$, bears the following $\ell$-read vector.
\begin{IEEEeqnarray*}{+rCl+x*}
\tr{\bfx} &=& (\mathcal{R}'_{1}, \ldots, \mathcal{R}'_{i}, \mathcal{R}'_{i}+1, \mathcal{R}'_{i+1}, \ldots, \mathcal{R}'_{n+\ell-2}).
\end{IEEEeqnarray*}

For the case of $\mathcal{R}'_{i+1}-\mathcal{R}'_{i} = -2$, the argument works similarly. Once $\tr{\bfx}$ is known, $\bfx$ can also uniquely recovered as suggested by \ref{prop::rec_rx}.
\end{IEEEproof}

\subsection{Upper Bound on the Size of Codes}\label{sec::upper}

This section establishes an upper bound on the cardinality of a single-deletion $\ell$-read code. We do so by limiting our focus to a subset of deletion patterns on an $\ell$-read vector, say $\tr{\bfx}$, which bear an intriguing connection with sticky deletions \cite{Levenshtein65, dolecekRepetitionErrorCorrecting2010, mahdavifarAsymptoticallyOptimalStickyinsertioncorrecting2017, wangCodesCorrectingLimitedMagnitude2023} on its respective binary vector, $\bfx$. {We consider a specific variant of a sticky deletion, defined below.}
\begin{defn}
	An \emph{$r$-sticky deletion}, for $r \geq 1$, is a deletion in a run of length at least $r$, in $\bfx$. \label{def::l-sticky}

  The error ball of a single $r$-sticky deletion for any $\bfu \in \Sigma_2^n$ is represented by 
  \begin{IEEEeqnarray*}{+rCl+x*}
	\sticky{\bfu}{r} &\triangleq& \big\lbrace (u_1u_2\ldots u_{i-1}u_{i+1}\ldots u_n) : i \in [n-r+1], \\
	&&  u_{i}=\cdots=u_{i+r-1} \big\rbrace. \label{eq::ell-sticky}
  \end{IEEEeqnarray*}
\end{defn}

Naturally, $\sticky{\bfu}{1}=\del{\bfu}$ and $|\sticky{\bfu}{r}| = \rho_{\geq r} (\bfu)$. 
As will be established shortly, such sticky deletions translate to specific deletion events on the respective $\ell$-read vectors, which we define formally as follows.

\begin{defn}
A \emph{$k$-restricted deletion} is a deletion that only deletes a symbol if it equals $0$ or $k$. \label{def::k-restr}
\end{defn}

The error ball of a $k$-restricted deletion for any $\bfu \in \Sigma_q^n$ can be expressed as
\begin{IEEEeqnarray*}{+rCl+x*}
\restr{\bfu}{k}  &\triangleq& \{ (u_1u_2\ldots u_{i-1}u_{i+1}\ldots u_n) : i \in [n-\ell+1], \\
&&  u_{i} \in \{0,k\} \}. \label{eq::ell-restr}
\end{IEEEeqnarray*}

The upcoming lemma explains the link between $\ell$-sticky deletions in binary vectors and $\ell$-restricted deletions in their respective $\ell$-read vectors.
\begin{lemma}
    For any $\bfx \in \Sigma_2^n$ with $\rho_{\geq \ell}(0^{\ell-1}\bfx 0^{\ell-1}) \geq 1$, it holds that 
    \begin{IEEEeqnarray*}{+rCl+x*}
    \restr{\tr{\bfx}}{\ell} &=& \{ \tr{\bfy} : 0^{\ell-1} \bfy 0^{\ell-1} \in \sticky{0^{\ell-1}\bfx 0^{\ell-1}}{\ell}  \}. \label{eq::equiv}
    \end{IEEEeqnarray*} \label{lem::l-sticky}
\end{lemma}
\begin{IEEEproof}
    Consider an $\mathcal{R}' \in \restr{\tr{\bfx}}{\ell}$, that arises from a deletion in $\tr{\bfx}$ at index $i$, and assume that $\tri{\bfx}{i}=0$, or equivalently, $\bfx_{i-\ell+1}^{i}=0^{\ell}$. For the case of $\tri{\bfx}{i}=\ell$, the proof follows similarly.

    If $\ell+1 \leq i \leq n$, we note that for $\bfy = (\bfx_1^{i-\ell} 0^{\ell-1} \bfx_{i+1}^{n}) \in \sticky{\bfx }{\ell}$, we get $\mathcal{R}'=\tr{\bfy}$, since $\bfx_1^{i-1}=\bfy_1^{i-1}$ ensures that $\tri{\bfy}{j}=\tri{\bfx}{j}$ for $j\in [i-1]$, while for $i\leq j \leq n+\ell-2$, we have $\tri{\bfy}{j}=\wt(\bfy^j_{j-\ell+1})=\wt(\bfx^{j+1}_{j-\ell+2})=\tri{\bfx}{j+1}$. 
    
    Next, consider $i\in [\ell]$. Since $\bfx_1^i=0^i$, it follows that for $\bfy = (0^{i-1} \bfx_{i+1}^n)$, $\mathcal{R}'=\tr{\bfy}$. 
    Observe that $(0^{\ell-1} \bfy) \in \sticky{0^{\ell-1}\bfx }{\ell}$.
    
    Finally when $i\geq n+1$, we can similarly argue that for $\bfy = (\bfx_1^{i-\ell}0^{\ell-(i-n)-1})$, which upholds $( \bfy 0^{\ell-1}) \in \sticky{\bfx 0^{\ell-1}}{\ell}$, $\mathcal{R}'=\tr{\bfy}$. In all of the aforementioned cases $0^{\ell-1}\bfy 0^{\ell-1} \in \sticky{0^{\ell-1}\bfx 0^{\ell-1}}{\ell}$ and since $\tr{\bfy}$ is fixed, $\bfy$ is clearly unique. 
    
    To prove the other direction, assume some $\bfx \in \Sigma_2^n$ and $\bfy \in \Sigma_2^{n-1}$ such that $\hat{\bfy} = 0^{\ell-1} \bfy 0^{\ell-1} \in \sticky{\hat{\bfx} = 0^{\ell-1}\bfx 0^{\ell-1}}{\ell}$. In particular, let the run (of length $\geq \ell$) in $\hat{\bfx}$ that suffers the $\ell$-sticky deletion start at index $j$, i.e., $\hat{\bfx}_j^{j+\ell-1}=a^{\ell}$ for some $a \in \Sigma_2$ and $\hat{\bfy} = \hat{\bfx}_1^{j-1} \hat{\bfx}_{j+1}^{n+2\ell-1}$. If $j=1$, then $\hat{\bfx}_1^{\ell} = \bfx_{-\ell+2}^1$, $\tri{\bfx}{1}=0$ and $\hat{\bfy} = 0^{\ell-1} \bfy 0^{\ell-1} =0^{\ell-1} \bfx_2^n 0^{\ell-1} $ directly imply that for all $p \in [n+\ell-2]$, $\tri{\bfy}{p} = \tri{\bfx}{p+1}$ and thereby $\tr{\bfy} \in \restr{\tr{\bfx}}{\ell}$.

    For the remaining cases, we have $j \in \{ \ell, \ldots, n+\ell-1\}$. Note that $\bfy_1^{j-\ell} =\hat{\bfy}_{\ell}^{j-1} = \hat{\bfx}_{\ell}^{j-1} =  \bfx_1^{j-\ell}$ in combination with $\bfy_{j-\ell+1}^{j-1}=\bfx_{j-\ell+1}^{j-1}=a^{\ell-1}$ establishes that for all $p\in [j-1]$, $\tri{\bfy}{p} = \tri{\bfx}{p}$. Evidently, $\tri{\bfx}{j} = \ell \cdot a \in \{0, \ell\}$. Next, from $\bfy_{j-\ell+1}^{n-1}0^{\ell-1}=\hat{\bfy}_j^{n+2\ell-3}=\hat{\bfx}_{j+1}^{n+2\ell-2}=\bfx_{j-\ell+2}^{n}0^{\ell-1}$, observe that for all $p \in \{j, \ldots, n+\ell-2\}$, $\tri{\bfy}{p} = \tri{\bfx}{p+1}$. Thus $\tr{\bfy} \in \restr{\tr{\bfx}}{\ell}$ and the statement follows.
\end{IEEEproof}

\begin{example}
    Recall $\bfx=(1,0,1,1,0,0)$ from Example~\ref{eg::read}, with the $3$-read vector $\tr{\bfx}=(1,1,2,2,2,1,0,0)$. 
    Consider $\mathcal{R}'=(1,1,2,2,2,1,0)$ that arises from deleting a $0$ in $\tr{\bfx}$, i.e., $\mathcal{R}' \in \restr{\tr{\bfx}}{\ell}$ where $\ell=3$. Observe that for $\bfy=(1,0,1,1,0)$, we have $0^{\ell-1}\bfy 0^{\ell-1} \in \sticky{0^{\ell-1}\bfx 0^{\ell-1}}{\ell}$ and $\tr{\bfy}=\mathcal{R}'$.
\end{example}

Thus for any $\bfx \in \Sigma_2^n$ with $\rho_{\geq \ell}(0^{\ell-1} \bfx 0^{\ell-1}) \geq 1$, it holds that $|\restr{\tr{\bfx}}{\ell}|=|\sticky{0^{\ell-1}\bfx 0^{\ell-1}}{\ell}| = \rho_{\geq \ell}(0^{\ell-1} \bfx 0^{\ell-1}) \geq \rho_{\geq \ell}(\bfx)$. 
The following corollary summarizes our strategy for bounding the size of any single-deletion-correcting read code.

\begin{corollary}
Any single-deletion $\ell$-read code is also a single-$\ell$-sticky-deletion-correcting code. \label{cor::sticky-read}
\end{corollary}
\begin{IEEEproof}
Based on Lemma~\ref{lem::l-sticky}, $\mathset*{\tr{\bfy}}{\bfy \in \sticky{\bfx}{\ell}} \subseteq \mathset*{\tr{\bfy}}{0^{\ell-1} \bfy 0^{\ell-1} \in \sticky{0^{\ell-1} \bfx 0^{\ell-1}}{\ell}} = \nolinebreak \restr{\tr{\bfx}}{\ell} \subseteq \del{\tr{\bfx}}$ for any~$\bfx\in \Sigma_2^n$. Then, for any two distinct codewords~$\bfc,\bfc'$ of a single-deletion $\ell$-read code we necessarily have $\mathset*{\tr{\bfy}}{\bfy \in \sticky{\bfc}{\ell}}\cap \mathset*{\tr{\bfy}}{\bfy \in \sticky{\bfc'}{\ell}} = \emptyset$, which in particular implies $\sticky{\bfc}{\ell}\cap \sticky{\bfc'}{\ell} = \emptyset$, i.e., the code is also single-$\ell$-sticky-deletion-correcting.
%
\end{IEEEproof}

Consequently, the cardinality of a single-deletion $\ell$-read code is bounded from above by the size of the largest code that corrects a single $\ell$-sticky deletion.



Now to establish an upper bound on the cardinality of a code that corrects a single $\ell$-sticky deletion. We first note that for a randomly chosen $\bfx \in \Sigma_2^n$, the expected 
value of $\rho_{\geq a}(\bfx)$ 
is given by $2^{-a} (n-a+2)$ (see Appendix). 
Also note that a change in any $x_i$ may increase or decrease $\rho_{\geq a}(\bfx)$ by at most~$1$. It then follows from McDiarmid's inequality \cite{Doo40} that for any $\epsilon>0$
\begin{IEEEeqnarray*}{+rCl+x*}
2^{-n} \abs*{\mathset*{\bfx\in\Sigma_2^n}{ \rho_{\geq a}(\bfx)  < \frac{n-a+2}{2^{a}} -\epsilon n}} \leq \exp\parenv*{- 2 \epsilon^2 n}.
\end{IEEEeqnarray*}
By choosing $\epsilon = 2^{-a-1}$, we obtain 
\begin{IEEEeqnarray*}{+rCl+x*}
\abs*{\mathset*{\bfx\in\Sigma_2^n}{ \rho_{\geq a}(\bfx)  < \frac{n-2a+4}{2^{a+1}}}} \leq 2^n \exp\parenv*{- \frac{n}{2^{2a+1}}}.\IEEEeqnarraynumspace
\label{eq::mcdiarmid}
\end{IEEEeqnarray*}

In the following analysis, assume that $\ell \geq 2$. We proceed to apply the generalized sphere-packing bound \cite{fazeliGeneralizedSpherePacking2015} to obtain an upper bound on the cardinality of the largest length-$n$ single $\ell$-sticky-deletion-correcting code, which we denote by $A(n,\ell)$. 


\begin{theorem}
    \cite{fazeliGeneralizedSpherePacking2015} 
    For an error channel that outputs any $\bfy \in B(\bfx)$ given an input $\bfx \in \cX$, construct a hypergraph $\cH(Y, \cE)$
    , whereby the vertex set $Y$ comprises all possible channel outputs while the hyperedge set $\cE = \{ B(\bfx) : \bfx \in \cX\}$. Assume an assignment of weights $w_{\bfy}$ for each $\bfy \in Y$, satisfying $w_{\bfy} \geq 0$ for all $y\in Y$, and $\sum_{\bfy \in E} w_{\bfy} \geq 1$ for all $E \in \cE$. Then, the size of any error-correcting code for this channel is upper bounded by $\sum_{\bfy \in Y} w_{\bfy}$.    \label{th::sphere}
\end{theorem}

The hypergraph for our single $\ell$-sticky deletion channel, say $\mathcal{H}(Y,\mathcal{E})$, constitutes the vertex set $Y = \Sigma_2^{n-1}$ and the hyperedge set $\mathcal{E} = \{\sticky{\bfx}{\ell} : \bfx \in \Sigma_2^n, \rho_{\geq \ell}(\bfx) \geq 1 \}$. The restriction on $\rho_{\geq \ell}(\bfx)$ simply serves to exclude the empty set. We choose the following weight assignment for the vertices in $Y$. For each $\bfy \in Y$,
\begin{IEEEeqnarray*}{+rCl+x*}
     w_{\bfy} \triangleq \begin{cases}
        \frac{1}{\rho_{\geq \ell}(\bfy)} & \text{ if } \rho_{\geq\ell}(\bfy) \geq 1, \\
        1 & \text{ if } \rho_{\geq\ell}(\bfy) = 0.
    \end{cases}
\end{IEEEeqnarray*}


While this clearly fulfills the positivity criterion of \cref{th::sphere}, observe that for all $\bfx \in \Sigma_2^n$ that fulfill either $\rho_{\geq\ell}(\bfx) \geq 2$ or $\rho_{\geq \ell}(\bfx)=\rho_{\geq \ell+1}(\bfx)=1$, it holds that $\rho_{\geq\ell}(\bfy) \geq 1$ for each $\bfy \in \sticky{\bfx}{\ell}$, since a single $\ell$-sticky deletion only shortens a single run by one. Thus $\sum_{\bfy \in \sticky{\bfx}{\ell}} w_{\bfy} = \sum_{\bfy \in \sticky{\bfx}{\ell}} \frac{1}{\rho_{\geq\ell}(\bfy)} \geq \sum_{\bfy \in \sticky{\bfx}{\ell}} \frac{1}{\rho_{\geq\ell}(\bfx)} = 1$. Similarly, when $\rho_{\geq\ell}(\bfx)=1$ while $\rho_{\geq\ell+1}(\bfx)=0$ (i.e., $\bfx$ has a unique longest run, of length $\ell$), it follows that $\sticky{\bfx}{\ell} = \{ \bfy \}$ wherein $\rho_{\geq \ell}(\bfy)=0$, leading us to $\sum_{\bfy \in \sticky{\bfx}{\ell}} w_{\bfy}=1$.

We now derive an upper bound on the cardinality~$A(n,\ell)$ of the largest length-$n$ single $\ell$-sticky-deletion-correcting code.
\begin{IEEEeqnarray*}{+rCl+x*}
    A(n,\ell) &\leq& \sum_{\bfy \in Y} w_{\bfy} = \abs*{\mathset*{\bfy\in \Sigma_2^{n-1}}{\rho_{\geq\ell}(\bfy) = 0}} \> +\\
    && \sum_{i=1}^{\floor{\frac{n-1}{\ell}}} \frac{1}{i} \abs*{\{\bfy \in \Sigma_2^{n-1} : \rho_{\geq \ell}(\bfy) = i \}} \\
    &<& \abs*{\mathset*{\bfy\in \Sigma_2^{n-1}}{\rho_{\geq\ell}(\bfy) < \frac{n-2\ell+3}{2^{\ell+1}}}} \>+ \\
    && \frac{1}{\ceilenv*{\frac{n-2\ell+3}{2^{\ell+1}}}} \sum_{i = \ceilenv*{\frac{n-2\ell+3}{2^{\ell+1}}}}^{\floor{\frac{n-1}{\ell}}} \abs*{\mathset*{\bfy\in \Sigma_2^{n-1}}{\rho_{\geq\ell}(\bfy) = i}} \\
    &\leq& 2^{n-1} \exp\parenv*{- \frac{n-1}{2^{2\ell+1}}} + \frac{2^{n+\ell}}{n-2\ell+3} \\
    &\leq& \frac{2^{n+\ell-1}}{n} \parenv*{2^{-\ell} n \exp\parenv*{- \frac{n-1}{2^{2\ell+1}}} + \frac{2n}{n-2\ell}}.
\end{IEEEeqnarray*}

We merge this with Corollary~\ref{cor::sticky-read} to get the following bound.

\begin{theorem}
    The redundancy of a single-deletion $\ell$-read code is bounded from below by
    $$\log n -\ell+1 - \log \parenv*{ \frac{2}{1-2\ell / n} + 2^{-\ell}n\exp\parenv*{-\frac{n-1}{2^{2\ell + 1}}} }.$$ \label{th::1del-bound}
\end{theorem}

\emph{Remark:} As mentioned earlier, for any $\bfx \in \Sigma_2^n$, the transverse-read vector as defined in \cite{cheeTransverseReadCodesDomainWall2023}, is a substring of $\tr[\ell]{\bfx}$ for certain choices of parameters. Consequently in these cases, \cref{lem::l-sticky} can be suitably modified to help establish a similar redundancy bound for the transverse-read channel.

\subsection{Single-Deletion $\ell$-Read Codes}

It is implied by \ref{prop::rec_rx} that protecting the first $n$ entries of the $\ell$-read vector taken modulo $2$ suffices to exactly recover the corresponding length-$n$ binary vector. This naturally leads us to the following single-deletion $\ell$-read code.

\begin{construction}
    \begin{IEEEeqnarray*}{+rCl+x*}
        \mathcal{C}(n, \ell, a)\! &= \! \{ \bfx\in \Sigma_2^n :  \! \sum_{i=1}^{n} i (\tri{\bfx}{i} \bmod 2)\! =a \hspace{-3mm} \pmod{n+1} \}.
    \end{IEEEeqnarray*} \label{cnstr::del}
    where $a \in \Sigma_{n+1}$.
\end{construction}

Note that for every $\bfu \in \Sigma_2^n$, there exists $\bfx \in \Sigma_2^n$ such that $\bfu=(\tri{\bfx}{1}, \ldots, \tri{\bfx}{n}) \bmod 2$. Thus by the pigeonhole principle, there exists $a \in \Sigma_{n+1}$ for which the redundancy required by the preceding construction is $\log_2 (n+1)$ bits, implying that 
\cref{cnstr::del} is optimal up to a constant. 

The accuracy of this construction is demonstrated below.


\begin{theorem}
    For all $a \in \Sigma_{n+1}$, the code $\mathcal{C}(n,\ell,a)$ is a single-deletion $\ell$-read code.
\end{theorem}
\begin{IEEEproof}
    Consider a vector $\mathcal{R}'$ resulting from a single deletion on the $\ell$-read vector of some $\bfx \in \mathcal{C}(n, \ell, a)$. Also, due to \cref{lem::del_g2}, we only consider the case when no pair of consecutive elements in $\mathcal{R}'$ have an absolute difference exceeding $1$, as the deletion is immediately correctable otherwise.
    
    Now consider a truncation of $\mathcal{R}'$, given by $\widetilde{\mathcal{R}}' = (\mathcal{R}'_1, \ldots, \mathcal{R}'_{n-1})$. Due to the VT constraint in \cref{cnstr::del} and the fact that $\widetilde{\mathcal{R}}' \bmod 2 \in \del{(\tri{\bfx}{1}, \ldots, \tri{\bfx}{n}) \bmod 2}$, one can uniquely recover $(\tri{\bfx}{1}, \ldots, \tri{\bfx}{n}) \bmod 2$, which suffices for the recovery of $\tr{\bfx}$ and thereby $\bfx$, 
    as indicated by  \ref{prop::rec_rx}.
\end{IEEEproof}

\section{Multiple reads}

DNA synthesis technologies typically generate multiple copies of each strand, while PCR amplification during sequencing tends to boost the number of copies even further, albeit introducing errors in the process \cite{churchNextGenerationDigitalInformation2012, goldmanPracticalHighCapacityLowMaintenance2013, organickRandomAccessLargescale2018, yazdiDNABasedStorageTrends2015}. Investigating how the availability of multiple noisy versions at the receiver might ease the reconstruction process, is thus a relevant and intriguing problem \cite{levenshteinEfficientReconstructionSequences2001, levenshteinEfficientReconstructionSequences2001a, chrisnataCorrecting2022, abu-siniLevenshteinReconstructionProblem2021, batuReconstructingStringsRandom2004, phuocphamSequenceReconstructionProblem2022, gabrysSequenceReconstructionDeletion2018, yehezkeallyReconstructionCodesDNA2020}. 

In one of the earliest works on reconstruction from multiple noisy sequences, Levenshtein \cite[Corollary 1]{levenshteinEfficientReconstructionSequences2001a} established that for any alphabet $\Sigma$ and two distinct vectors $\bfu, \bfv \in \Sigma^n$, it holds that $|\del{\bfu} \cap \del{\bfv}| \leq 2$. Now given any two distinct binary vectors $\bfx, \bfy \in \Sigma_2^n$, we may replace $\bfu$ and $\bfv$ with $\tr{\bfx}$ and $\tr{\bfy}$ respectively, and thereby conclude that three noisy versions of an $\ell$-read vector are sufficient to uniquely determine the channel input. To examine if and how the intrinsic characteristics of $\ell$-read vectors might allow for more efficient reconstruction strategies, we endeavor to assess in this section, how the availability of two distinct noisy $\ell$-read vectors might lower the redundancy required to correct a single deletion. Equivalently, we are interested in the conditions under which two distinct binary vectors $\bfx, \bfy \in \Sigma_2^n$ have $\ell$-read vectors such that $|\del{\tr{\bfx}} \cap \del{\tr{\bfy}}| = 2$. 
 To accomplish this, we employ the following proposition from \cite{caiCodingSequenceReconstruction2022}.

\begin{defn}
    \cite{caiCodingSequenceReconstruction2022}
    Two words $\bfu$ and $\bfv$ of length $n$ are said to be \emph{confusable} if there exist subwords $\bfa$, $\bfb$ and $\bfc$ such that
    \begin{itemize}
        \item $\bfu=\bfa \bfc \bfb$ and $\bfv=\bfa \overline{\bfc} \bfb$ with $|\bfc|=|\overline{\bfc}| \geq 2$;
        \item $\{\bfc, \overline{\bfc}\}=\{ \alpha \beta \alpha \beta \ldots \alpha \beta, \beta \alpha \beta \alpha \ldots \beta \alpha \}$,
    \end{itemize}
    for some $\alpha,\beta\in \Sigma_q$.
\end{defn}

\emph{Remark:} Such vectors are named Type-A confusable in \cite{caiCodingSequenceReconstruction2022}.

\begin{propT}
\cite{caiCodingSequenceReconstruction2022} \label{propT::type-a} 
For any two distinct words $\bfu, \bfv \in \Sigma_q^n$, it holds that if $d_H(\bfu, \bfv)\geq 2$, we have ${|\del{\bfu} \cap \del{\bfv}| = 2}$ if and only if $\bfu$ and $\bfv$ are confusable.
\end{propT}

Thus, we seek to ascertain when and how two binary vectors might possess confusable $\ell$-read vectors. 

\begin{lemma}
    When $\ell\geq 2$
    , there exist no two distinct $\bfx, \bfy \in \Sigma_2^n$ such that $\tr{\bfx}$ and $\tr{\bfy}$ are confusable. \label{lem::type-a-read}
\end{lemma}

\begin{IEEEproof}
    We prove this by contradiction, i.e., we proceed by assuming the existence of two distinct $\bfx,\bfy \in \Sigma_2^n$ such that their respective read vectors are confusable. 
    
    Let $i$ refer to the first index where $\tr{\bfx}$ and $\tr{\bfy}$ disagree, implying that $\bfx_1^{i-1}=\bfy_1^{i-1}$. Now assume w.l.o.g. that $(x_i,y_i)=(0,1)$, causing $\tri{\bfy}{i}=\tri{\bfx}{i}+1$. For simplicity of exposition, we let $\tri{\bfx}{i}$ and $\tri{\bfy}{i}$ be denoted by $\alpha$ and $\beta$ respectively, where $\beta = \alpha+1$. By virtue of the confusability of the read vectors, we know that for some $m>0$, $(\tri{\bfx}{i}, \ldots, \tri{\bfx}{i+2m-1})= (\alpha \beta)^m$ while $(\tri{\bfy}{i}, \ldots, \tri{\bfy}{i+2m-1})= (\beta \alpha)^m$. Now the fact that $\tri{\bfx}{i+1}-\tri{\bfx}{i}=\tri{\bfy}{i}-\tri{\bfy}{i+1}=1$ necessitates $(x_{i+1}, x_{i-\ell+1})=(y_{i-\ell+1}, y_{i+1})=(1,0)$. However for $\ell \geq 2$, the requirement $x_{i-\ell+1} \neq y_{i-\ell+1}$ contradicts $\bfx_1^{i-1}=\bfy_1^{i-1}$. Thus, $\bfx$ and $\bfy$ do not exist.
\end{IEEEproof}

For $\ell \geq 2$, the above lemma implies the following outcome on the redundancy needed to uniquely recover a binary vector from two distinct erroneous copies of its $\ell$-read vector.

\begin{lemma}
    When $\ell\geq 2$
    , for any two distinct $\bfx, \bfy \in \Sigma_2^n$, $|\del{\tr{\bfx}} \cap \del{\tr{\bfy}}| \leq 1$. \label{lem::del2_max_overlap}
\end{lemma}

\begin{IEEEproof}
    We know from \cite[Corollary 1]{levenshteinEfficientReconstructionSequences2001a} that $|\del{\tr{\bfx}} \cap \del{\tr{\bfy}}| \leq 2$ since distinct binary vectors have distinct $\ell$ read vectors, as suggested by \ref{prop::rec_rx}. Additionally \cite[Lemma~1]{banerjeeErrorCorrectingCodesNanopore2023} asserts that for $\ell \geq 2$, it holds that $d_H(\tr{\bfx}, \tr{\bfy}) \geq 2$ for distinct $\bfx$ and $\bfy$. Upon combining these facts with Proposition~\ref{propT::type-a} and \cref{lem::type-a-read}, we arrive at the statement of the lemma.
\end{IEEEproof}

We can thus infer the following on the redundancy required for reconstruction with two noisy read vectors.

\begin{corollary}
    For any $\ell \geq 2$, $\bfx \in \Sigma_2^n$ and given any two distinct noisy read vectors $\cR', \mathfrak{R}' \in \del{\tr{\bfx}}$, $\tr{\bfx}$ can be uniquely recovered.
\end{corollary}

One possible method to accomplish reconstruction with two corrupted $\ell$-read vectors is outlined in Algorithm~\ref{alg::rec} 
and its correctness is proved in the next lemma.

\begin{lemma}
    For any $\ell \geq 2$ and $\bfx \in \Sigma_2^n$ such that $\del{\tr{\bfx}} \geq 2$, given any two distinct vectors in $\del{\tr{\bfx}}$, Algorithm~\ref{alg::rec} returns $\tr{\bfx}$. \label{lem::rec}
\end{lemma}
\begin{IEEEproof}
    Let the two noisy reads be denoted by $\mathcal{R}'$ and $\mathfrak{R}'$ respectively. By virtue of \cref{lem::del_g2}, we deem it sufficient to study the case wherein neither of these vectors has a pair of consecutive elements with an absolute difference exceeding $1$. 
    
    Let $i$ and $j$ denote the first and last indices at which $\mathcal{R}'$ and $\mathfrak{R}'$ disagree. Of course, $i=j$ when $d_H(\mathcal{R}',\mathfrak{R}')=1$. 
    Also assume that $\mathcal{R}'$ and $\mathfrak{R}'$ arise from a deletion on $\tr{\bfx}$ at indices $a$ and $b$ respectively. Evidently, $\{a,b \} = \{ i,j+1 \}$.

    Now consider the vectors
    \begin{IEEEeqnarray*}{+rCl+x*}
        \widehat{\mathcal{R}}(\bfx) &= (\mathcal{R}'_1, \ldots, \mathcal{R}'_{i-1}, \mathfrak{R}'_i, \mathcal{R}'_{i}, \ldots, \mathcal{R}'_{n+\ell-2}), \\
        \widetilde{\mathcal{R}}(\bfx) &= (\mathcal{R}'_1, \ldots, \mathcal{R}'_{j}, \mathfrak{R}'_j, \mathcal{R}'_{j+1}, \ldots, \mathcal{R}'_{n+\ell-2}).
    \end{IEEEeqnarray*}

    These are clearly distinct and depending on whether $a=i$ or $a=j+1$, $\tr{\bfx}$ is either equal to $\widehat{\mathcal{R}}(\bfx)$ or $\widetilde{\mathcal{R}}(\bfx)$. Next, observe that $\widehat{\mathcal{R}}(\bfx)$ and $\widetilde{\mathcal{R}}(\bfx)$ cannot be legitimate read vectors according to \cite[Proposition~1]{banerjeeErrorCorrectingCodesNanopore2023}, simultaneously, as \cref{lem::del2_max_overlap} would be otherwise contradicted. Thus, Algorithm~\ref{alg::rec} chooses one of $\widehat{\mathcal{R}}(\bfx)$ and $\widetilde{\mathcal{R}}(\bfx)$, as advised by  \cite[Proposition~1]{banerjeeErrorCorrectingCodesNanopore2023}.    
\end{IEEEproof}

\begin{algorithm}[t]
	\label{alg::rec}
	\DontPrintSemicolon
	\KwInput{$n$, $\ell$, set $\{ \mathcal{R}', \mathfrak{R}' \} \subseteq \del{\tr{\bfx}}$ for some $\bfx \in \Sigma_2^n$}
	\KwOutput{$\tr{\bfx}$}
	\Init{}{
 
        Let $i$ and $j$ be the first and last indices at which $\mathcal{R}'$ and $\mathfrak{R}'$ disagree.

        $\widehat{\mathcal{R}}(\bfx) \leftarrow (\mathcal{R}'_1, \ldots, \mathcal{R}'_{i-1}, \mathfrak{R}'_i, \mathcal{R}'_{i}, \ldots, \mathcal{R}'_{n+\ell-2})$;

        $\widetilde{\mathcal{R}}(\bfx) \leftarrow (\mathcal{R}'_1, \ldots, \mathcal{R}'_{j}, \mathfrak{R}'_j, \mathcal{R}'_{j+1}, \ldots, \mathcal{R}'_{n+\ell-2})$.

	}
    \If{$\widehat{\mathcal{R}}(\bfx)$ is the $\ell$-read vector\footnotemark  of any vector in $\Sigma_2^n$ }
    {
            $\tr{\bfx} \leftarrow \widehat{\mathcal{R}}(\bfx)$.
    }
    \Else
	{
    	  $\tr{\bfx} \leftarrow \widetilde{\mathcal{R}}(\bfx)$.
	}
	\caption{Reconstruct} 
\end{algorithm}

\footnotetext{For more details, and a description of an efficient verification procedure, we refer the reader to  \cite{banerjeeErrorCorrectingCodesNanopore2023}.}

\emph{Remark:} Prior work \cite{Levenshtein65, sloaneSingleDeletionCorrectingCodes2002, chrisnataCorrecting2022} states that for the standard single deletion channel, i.e., $\ell = 1$, the required redundancy decreases gracefully from $\log_2 n + O(1)$ to $\log_2 \log_2 n - O(1)$, given one and two distinct erroneous copies of an $\ell$-read vector, respectively. It is thus {somewhat} surprising to learn that when $\ell \geq 2$, the minimal redundancy cost remains the same for one received sequence, while for two noisy copies, it instantly drops to $0$. {This behavior can perhaps be attributed to the close connection between certain deletions in the read vector and sticky deletions in the original sequence, established in Lemma~\ref{lem::l-sticky}}{, as it is known that two distinct erroneous reads are sufficient for reconstruction from a single sticky deletion (an equivalent result for sticky insertions appears, e.g., in \cite{yehezkeallyReconstructionCodesDNA2020}).}


\section{Conclusion}

This work investigates how the inherent redundancy imbued by the physical aspects of a nanopore sequencer into its reads might help achieve more efficient deletion-correcting codes. To this end, we found that for the simplified model of nanopore sequencing adopted here, the minimal redundancy required to correct a single deletion reduces remarkably when the receiver is provided two distinct erroneous received sequences, as opposed to just one. This raises further questions regarding the case of multiple deletions and more received sequences, which we plan to explore in future work.

\printbibliography

\appendix

\subsection{Expected number of runs exceeding a given length}

\begin{claim}
    For any $n>0$ and $a \in [n]$, the expected number of runs with length greater than or equal to $a$, in a binary vector of length $n$ wherein each bit is chosen uniformly, is given by $E[\rho_{\geq a}(\bfx)]=2^{-a}(n-a+2)$.
\end{claim}
\begin{IEEEproof}
    For some $\bfx \in \Sigma_2^n$, let $\mathbbm{1}_{i,a}(\bfx)$ refer to the indicator function that evaluates to $1$ if a run of length greater than or equal to $a$ begins exactly at the $i$th index of $\bfx$, and $0$ otherwise. Hence, 
    \begin{IEEEeqnarray*}{+rCl+x*}
        E[\rho_{\geq a}(\bfx)] &=& \sum_{i = 1}^{n-a+1} E[\mathbbm{1}_{i,a}(\bfx)] \\
        &=& P(\bfx_1^a \in \{0^a, 1^a\}) \\
        &&+ \sum_{i = 2}^{n-a+1} P(\bfx_{i - 1}^{i + a - 1} \in \{10^a, 01^a\}) \\
        &=& 2^{-a}(n - a + 2). \\[-\normalbaselineskip] &&&\IEEEQEDhere
    \end{IEEEeqnarray*}
\end{IEEEproof}

\IEEEtriggeratref{4}

\end{document}